# Thermoelectric properties and electronic structure of Cr(Mo,V)N$_x$ thin films studied by synchrotron and lab-based X-ray spectroscopy

Susmita Chowdhury*, Victor Hjort, Rui Shu, Grzegorz Greczynski, Arnaud le Febvrier, Per Eklund, and Martin Magnuson

*Thin Film Physics Division, Department of Physics, Chemistry and Biology (IFM), Linköping University, Linköping SE-581 83, Sweden*

**Abstract**

Chromium-based nitrides are used in hard, resilient coatings, and show promise for thermoelectric applications due to their combination of structural, thermal, and electronic properties. Here, we investigated the electronic structures and chemical bonding correlated to the thermoelectric properties of epitaxially grown chromium-based multicomponent nitride Cr(Mo,V)N$_x$ thin films. Due to minuscule N vacancies, finite population of Cr *3d* and N *2p* states appear at the Fermi level and diminishes the band opening for Cr$_{0.51}$N$_{0.49}$. Incorporating holes by alloying V in N deficient CrN matrix results in enhanced thermoelectric power factor with marginal change in the charge transfer of Cr to N compared to Cr$_{0.51}$N$_{0.49}$. Further alloying Mo isoelectronic to Cr increases the density of states across the Fermi level due to hybridization of the (Cr, V) *3d* and Mo *4d*-N *2p* states in Cr(Mo,V)N$_x$. The hybridization effect with reduced N *2p* states off from stoichiometry drives the system towards metal like electrical resistivity and reduction in Seebeck coefficient compensating the overall power factor still comparable to Cr$_{0.51}$N$_{0.49}$. The N deficiency also depicts a critical role in reduction of the charge transfer from metal to N site. The present work envisages ways for enhancing thermoelectric properties through electronic band engineering by alloying and competing effects of N vacancies.

# 1. Introduction

Chromium nitride (CrN) is important in a range of existing and prospective applications due to its combination of structural, mechanical, magnetic, and electronic properties.[1] CrN undergoes a magnetostructural transition from a paramagnetic B1 NaCl (rocksalt) type of structure (Fm$\bar{3}$m) at room temperature to an antiferromagnetic orthorhombic structure (Pnma) at around 260 – 285 K temperature regime[2–4] accompanied by a debatable semiconducting/insulating-to-metallic transition.[4–6] Given that CrN is based on abundant and relatively cheap raw materials, it is promising for large-scale production, and CrN-based thin films are long in use in hard-coating applications.[7–9]

Conventional thermoelectric materials such as tellurides and antimonides are used to harvest thermoelectric power from waste energy,[10–12] but are limited because of scarcity and toxicity of the constituent elements.[13,14] An emerging class of alternative materials are transition metal nitrides, in particular those based on ScN and CrN.[15,16] With a thermoelectric power factor of 1.5-5 mW m$^{-1}$ K$^{-2}$ and relatively low thermal conductivity of 2-4 W m$^{-1}$K$^{-1}$ due to strong spin-lattice coupling,[17,18] CrN is comparable to the conventional Bi$_2$Te$_3$ and PbTe.[1,19] Thermoelectric properties, *i.e.*, Seebeck coefficient ($S$), electrical conductivity ($\sigma$) and thermal conductivity ($\kappa$) are strongly coupled and hence hard to optimize. Generally, reducing dimensionality of materials, inducing defects (hole/electron doping or grain boundaries) by alloying, or tuning the atomic masses of alloying elements, are alternatives to improve the thermoelectric properties of the parent compound.[20–23] The generalized findings are valid for chromium based multicomponent nitrides compared to binary CrN, manifesting higher hardness,[24] thermal stability[25] and enhanced thermoelectric performances.[26]

However, the thermoelectric performances are strongly correlated to the electronic structure of the final compound. Theoretical band structure calculations reveal a local and sharp increase of the density of states (DOS) near the Fermi level (E$_F$) for any thermoelectric materials.[27] Hence, to gain improved power factor (S$^2\sigma$), the DOS should be as large as possible around E$_F$ for increased $\sigma$ and as asymmetric as possible to achieve best S.[28] It occurs due to the resonance of either the valence or conduction band of the host semiconductor with an energy level of the localized atom in the compound and can to a first approximation be explained by the Mott equation.[21,27]

$$S = \frac{\pi^2}{3} \frac{k_B^2 T}{q} \left[\frac{d\{\ln \sigma(E)\}}{dE}\right]_{E=E_F} \quad \ldots\ldots\ldots (1)$$

where, k$_B$ = Boltzmann constant, q = charge, and T = absolute temperature. The expression for $\sigma$(E) is given by,[21]

$$\sigma(E) = n(E)q\mu(E) \quad \ldots\ldots\ldots (2)$$

Here, n(E) = charge carrier density and µ(E) = mobility of the charge carrier. The charge carrier density is related to the DOS g(E) and Fermi function f(E) as,[21]

$$n(E) = g(E)f(E) \quad \ldots\ldots\ldots (3)$$

Furthermore, theoretical approaches for strongly-correlated-electron systems like CrN indicate that induced defects (metal/N vacancies) lead to an increase in the DOS along with shift in the E$_F$ affecting the band opening.[29] This in turn affects the thermoelectric power factor S$^2\sigma$.[29,30] Among different alloying elements (e.g., Sc, V, Al, Mo, W) in CrN, V alloying resulted in enhanced thermoelectric properties of $Cr_{1-x}V_xN$ in both bulk and thin films, due to increased hole concentration.[31,32] While alloying heavier Mo atoms, isoelectronic to Cr, would not alter the electronic properties substantially but rather affect thermal transport properties by increased phonon scattering.[33]

Consequently, in the present study attempts were made to synthesize epitaxial CrN, $Cr_{1-x}V_xN$ and Cr(Mo,V)N thin films. To the best of our knowledge, no literature on such complex chromium-based multicomponent nitride $Cr(Mo,V)N_x$ has been reported so far. Since probing electronic structure in any thermoelectric material is primordial for further improvement of transport properties, below the $E_F$, the DOS were probed by synchrotron-based resonant inelastic X-ray scattering (RIXS) (*partial*-DOS) complementary to lab-based valence band spectroscopy (VBS) (*total*-DOS). RIXS study on correlated chromium nitride-based system is non-existent, maybe due to poor energy resolution in earlier times.[34] The valence band below the $E_F$ was also studied using lab-based X-ray photoelectron spectroscopy (XPS) which is highly surface-sensitive with overlapping spectral features. In addition, above the $E_F$ synchrotron-based X-ray absorption (XAS) measurements were performed probing the unoccupied states in the conduction band. To investigate the thermoelectric properties, electrical resistivity and Seebeck coefficient of the samples were measured at room temperature. Thus, the structural, electronic, and thermoelectric correlations were systematically and quantitatively studied for CrN and $Cr_{1-x}V_xN$ thin films. Later, we qualitatively delve into the more complex systems of chromium-based multicomponent nitride $Cr(Mo,V)N_x$ thin films.

## 2. Experimental details

CrN, $Cr_{1-x}V_xN$ and a series of $Cr(Mo,V)N_x$ thin film samples were deposited on single-side polished c-plane sapphire (0001) substrates using reactive dc magnetron sputtering in an ultrahigh-vacuum deposition system described elsewhere.[35] The substrates were left electrically floating at a deposition temperature of 600°C. Depositions were performed using three magnetrons, each with > 99.7 % pure metal targets, and in an atmosphere of Ar and $N_2$. The gas composition was fixed at 40 % Ar and 60 % $N_2$. The CrN reference was deposited at 0.32 Pa and 22 sccm Ar, while the rest of the samples were deposited at 0.40 Pa and 28 sccm Ar, due to difficulty in sustaining the plasma of all three targets ignited at lower gas flow. More detailed description of depositions and more in-depth characterization can be found elsewhere.[36]

Rutherford backscattering (RBS) measurements were performed at Uppsala University using 2 MeV $^4He^+$ ion beam.[37] Backscattered ions were detected at a scattering angle of 170°. Channeling effects in the substrates and samples were minimized by adjusting the equilibrium incidence angle to 5° with respect to the surface normal and perform multiple-small-random-angular movements within a range of 2° during data acquisition. Atomic concentrations were extracted from the spectra using the SIMNRA simulation program.[38]

The X-Ray Diffraction (XRD) measurements were performed in Bragg-Brentano-mode (θ-2θ) using a PANalytical X'Pert Pro diffractometer system, with a Cu-Kα source operated at 45 kV and 40 mA. The incident optics was a Bragg-Brentano module with 0.5° divergence slit and a 0.5° anti-scatter slit, while the diffracted optics included a 5.0 mm anti-scatter slit, a 0.04 rad Soller slit, a Ni-filter, and an X'Celerator detector. Detailed structural analysis using pole figures and transmission electron microscopy is described elsewhere.[36]

The soft X-ray absorption near-edge structure (XANES) at Cr *2p*, N *1s*, Mo *3p* and V *2p* were measured at the SPECIES beamline equipped with an elliptically polarizing undulator (EPU61) and a plane grating monochromator (PGM), at the MAX IV Laboratory, Lund, Sweden. The XANES spectra were measured at 20° grazing incidence with 0.1 eV resolution using total electron yield (TEY) and total fluorescence yield (TFY), simultaneously. The combination of drain current and NEXAFS detector enabled to acquire both surface and bulk sensitive information simultaneously. For normalization of the data, a 4 µm thick Au reference foil was scanned in the same energy range as the samples over each absorption edge.

The RIXS spectra were also measured at SPECIES beamline with a high-resolution Rowland-mount grazing-incidence grating spectrometer[39,40] with a two-dimensional multichannel detector with a resistive anode readout. The Cr *2p* and N *1s* RIXS spectra were recorded using a spherical grating with 1200 lines/mm of 5 m radius in the first order of diffraction. During the Cr *2p* and N *1s* RIXS measurements, the energy resolutions of the beamline monochromator were 0.45, and 0.2 eV, respectively. The spectrometer resolutions were 0.4 for Cr *2p* and 0.3 eV for N *1s* spectra. All measurements were performed with a base pressure lower than $6.7 \times 10^{-7}$ Pa. In order to minimize self-absorption effects[41] the angle of incidence was 20° from the surface plane during the emission measurements. The x-ray photons were detected parallel to the polarization vector of the incoming beam to minimize elastic scattering.

The hard X-ray XANES and EXAFS measurements were performed at Cr *K*-edge in fluorescence mode at the BALDER beamline[42] at MAX IV. For reference and energy calibration, both XANES and EXAFS were performed on a 5 µm thick Cr foil (*K*-edge at 5989 eV) in transmission mode. The energy scans were done using a Si (111) double crystal monochromator and either a 7-element silicon drift detector (X-PIPS, from Mirion Technologies for fluorescence signal) or ionization gas detector filled with Ar, $N_2$ and He gases (for transmission signal) were used to measure the signals. A hexapod sample holder was used to mount the samples which were placed at an incidence and exit angle of 45° from the source and the detector, respectively in fluorescence mode. However, in the transmission mode the Cr foil was fixed at 90° incidence angle. For both XANES and EXAFS, the energy scans were repeated three times for each sample, at an energy interval of 0.25 eV and 0.5 eV with an integration time of 0.02 s. For fitting the EXAFS spectra, scattering lengths of the photoelectron and the phase shift were calculated using the FEFF9 code[43,44] considering the body centered cubic (bcc) and NaCl rocksalt-type B1 structure of Cr (COD-ID 5000220) and CrN (COD-ID 1010974), respectively. The data processing was done using the Visual Processing in EXAFS Researches (VIPER) software[45] and three scans for each spectrum from the 7 channels of the detector were analyzed and summed in the software to generate the final spectrum for each sample. A modified Victoreen polynomial function was used for the pre-edge normalization and a smoothing spline function was used for the post-edge background correction of the XAFS spectra. The $k^2$-weighted back Fourier Transform (FT) spectra were fitted in the range of 0-13.5 Å$^{-1}$ after extracting from the forward FT spectra within R-φ = 1-3 Å.

Core-level XPS measurements were performed in an Axis Ultra DLD instrument, Kratos Analytical (UK), using monochromatized Al-Kα radiation (1486.6 eV). The base pressure during analysis was ~$1.3 \times 10^{-7}$ Pa. Prior to measurements, samples were sputter etched for 10 minutes using 0.5 keV Ar$^+$ ions incident an angle of 70° from the surface normal. The area affected by the Ar$^+$ beam was 3×3 mm$^2$, while the analysis area was 0.3×0.7 mm$^2$ (centered in the middle of the etched crater). All spectra are charge referenced by setting the low energy

DOS cut-off at 0 eV. For $Cr_{0.44}Mo_{0.08}V_{0.06}N_{0.42}$ and $Cr_{0.45}Mo_{0.09}V_{0.07}N_{0.40}$ sample, an electron flood gun was used to neutralize the charge accumulation on the sample surface due to low electrical conductivity. Complementary to *p*-DOS, the total DOS was probed using valence band spectroscopy (VBS).

The sheet resistances of the samples were measured using a standard four-point-probe set-up (Jandel Model RM3000) at room temperature (~300 K) with equidistant probes with spacing of 1 mm each and a tip radius of 100 µm. The electrical resistivities ($\rho_{el}$) of the samples were then calculated taking into consideration the film thickness[46] as deduced from the XRR measurements (Table S1 in supplementary information).[47] The Seebeck coefficients were also measured at room temperature using a home-built thermoelectric measurement setup.[48] The set-up was equipped with two Peltier heat sources for creating a temperature gradient in the sample and two K-type thermocouples for measuring the temperature. The two electrodes are made of Cu and are in contact with the sample in an area of approximately 9×1 mm$^2$ in which the K-type thermocouples are present.

## 3. Results and Discussion

### 3.1 Compositional Analysis

**Table I**. Details of metal film composition obtained from RBS. The lattice parameters were calculated from the 111 peaks of each sample. The absorption edges around the Cr *K*-edge spectra are also listed.

| Samples | Film composition | | | | | Lattice Parameter (±0.004) (Å) | Absorption Edge (eV) |
|---|---|---|---|---|---|---|---|
| | RBS (% of Me) (±1%) | | | | | | |
| | Cr | Mo | V | N | N/Me ratio | | |
| $Cr_{0.51}N_{0.49}$ | 51 | - | - | 49 | 0.96 | 4.159 | 5997.9 |
| $Cr_{0.50}V_{0.03}N_{0.47}$ | 50 | - | 3 | 47 | 0.89 | 4.148 | 5997.7 |
| $Cr_{0.44}Mo_{0.08}V_{0.06}N_{0.42}$ | 44 | 8 | 6 | 42 | 0.72 | 4.072 | 5997 |
| $Cr_{0.45}Mo_{0.09}V_{0.07}N_{0.40}$ | 44.5 | 8.5 | 7 | 40 | 0.67 | 4.061 | 5997 |
| $Cr_{0.44}Mo_{0.09}V_{0.08}N_{0.39}$ | 44 | 9 | 8 | 39 | 0.64 | 4.030 | 5996.3 |

Table I shows the compositional analysis of $Cr_{0.51}N_{0.49}$, $Cr_{0.50}V_{0.03}N_{0.47}$ and $Cr(Mo,V)N_x$ samples measured by RBS. The composition of the films deduced by fitting revealed substoichiometry in nitrogen. The sample $Cr_{0.51}N_{0.49}$ is close to stoichiometry (N/Me ratio = 0.96) while $Cr_{0.50}V_{0.03}N_{0.47}$ and $Cr(Mo,V)N_x$ are substoichiometric with composition from $Cr_{0.50}V_{0.03}N_{0.47}$ to $Cr_{0.44}Mo_{0.09}V_{0.08}N_{0.39}$ for the lowest N containing film. The sample $Cr_{0.51}N_{0.49}$ and $Cr_{0.50}V_{0.03}N_{0.47}$ are used as reference while the other three samples contain around 8 - 9% of Mo in the metal site and an increase in the V content from 6 to 8%. All the samples contain negligible amount of oxygen which is less than the detection limit (~1 atomic %) of the instrument. A plausible explanation for the higher degree of N deficiency within the $Cr(Mo,V)N_x$ series compared to $Cr_{0.51}N_{0.49}$ and $Cr_{0.50}V_{0.03}N_{0.47}$ is explained below (see section 3.2.1).

### 3.2 Structural Analysis

*3.2.1 X-ray diffraction*

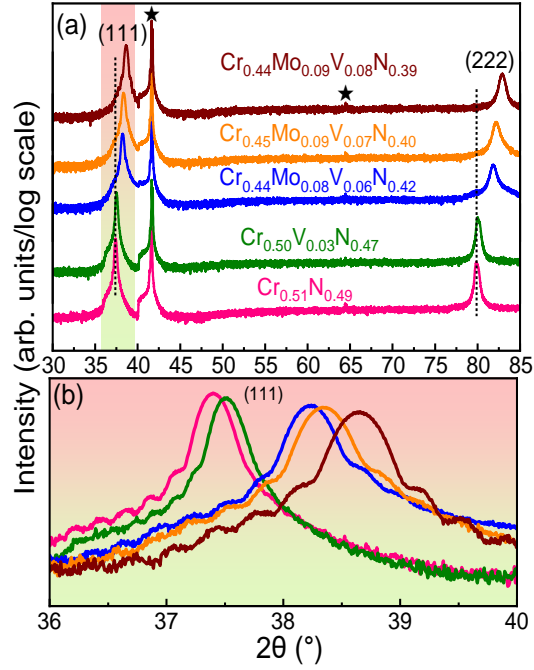

**Figure 1:** θ-2θ X-ray diffraction patterns where the stars represent the corresponding substrate peaks (a), and magnified view around the 111 peak (b) of $Cr_{0.51}N_{0.49}$, $Cr_{0.50}V_{0.03}N_{0.47}$ and $Cr(Mo,V)N_x$ thin films deposited on c-plane (0001) sapphire substrates.

Figure 1(a) shows θ-2θ XRD patterns of $Cr_{0.51}N_{0.49}$, $Cr_{0.50}V_{0.03}N_{0.47}$ and $Cr(Mo,V)N_x$ epitaxial thin films, described in more details elsewhere.[36] For all samples, the (0001)-oriented $Al_2O_3$ substrate provides a template for twin-domain epitaxial cubic growth of the thin films along [111] direction.[49] With increase in alloying-element concentration of V and incorporation of Mo in the CrN matrix, the lattice parameter gradually decreases as evident from the shift of the 111 and 222 peaks to higher diffraction angles. Up to the highest alloying element concentration, a single-phase cubic NaCl structure is retained, with no indication of any secondary phases. The 111 peaks of each sample were fitted using pseudo-Voigt functions and the lattice parameters determined from the 111 peak positions are listed in Table I. It can be inferred from the present study that addition of V increases the solid solubility limit of $Cr(Mo,V)N_x$ preventing phase segregation of $Mo_2N$ as observed by Quintela *et al* in bulk $Cr_{1-x}Mo_xN$ (at x ≥ 0.025).[33] For $Cr_{0.51}N_{0.49}$, the lattice parameter corroborates well with literature values around 4.15-4.18 Å.[49] Note the presence of Laue oscillations (see Figure 1(b)) in all the samples, indicating high crystallinity.

The atomic radii of Cr, V, Mo, and N are 140, 135, 145 and 65 pm, respectively. It is known that with nitridation of Cr, phase transition from Cr (bcc) → β-$Cr_2N$ (hcp) → CrN (rocksalt) occurs.[50] For stoichiometric CrN, the N atoms occupy 100% of the interstitial octahedral sites of the metal lattice. Earlier studies show alloying with V or Mo results in an increase in the lattice parameter provided the Cr atoms are substituted by the metal atoms in bulk CrN.[31,33] Nevertheless, our RBS results confirm the presence of N vacancies in all the samples. Thus, decrease in the lattice parameter stems from the volume contraction caused by the insufficient N occupancy. Similar observations have been reported for N deficient CrN thin films.[50] Although the metal content is considerably more than 50% (and N content is reduced from 47 to 39%) of the lattice site, the B1 rocksalt structure is still retained, instead of transformation to hcp $Cr_2N$. Within the $Cr(Mo,V)N_x$ samples, the reduction in the lattice parameter follows the trend of B1 MoN$_x$ (4.20 – 4.27 Å)/hcp MoN (a = 5.72 Å, c = 5.60 Å) → fcc Mo$_2$N (4.16 – 4.19 Å)[51] with reduction in the N content. Since, the lattice parameters of fcc Mo$_2$N and rocksalt CrN matches closely, addition of Mo leads to half occupancy of N in the non-metal site (approaching Mo$_2$N) resulting in more N deficiency in Cr(Mo,V)N$_x$ samples compared to Cr$_{0.51}$N$_{0.49}$ and Cr$_{0.50}$V$_{0.03}$N$_{0.47}$.

*3.2.2 Extended X-ray Fine Structure (EXAFS)*

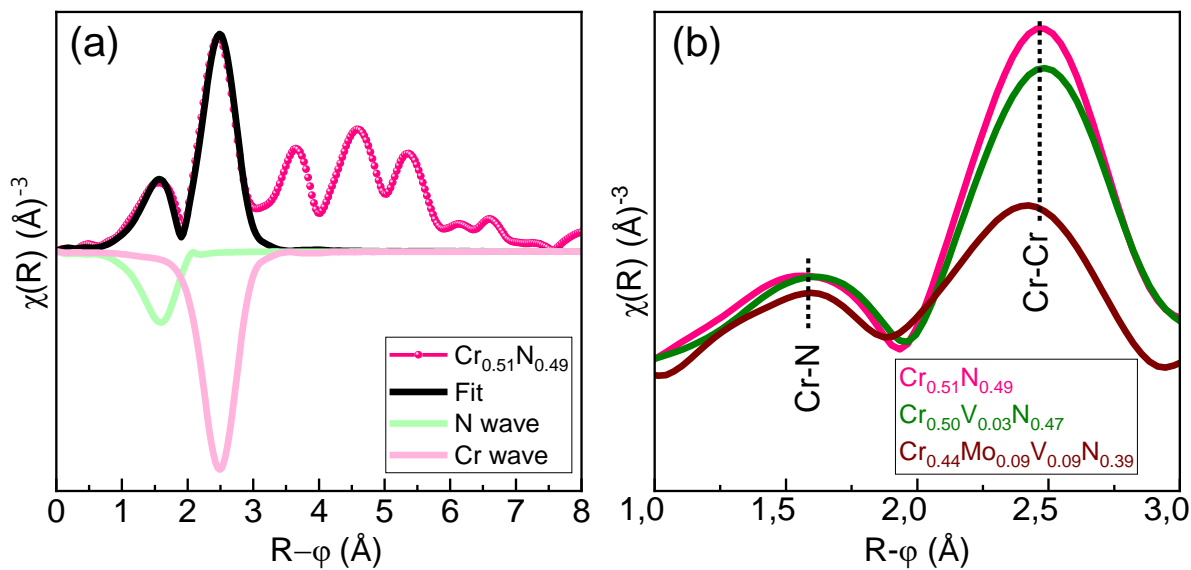

**Figure 2:** Fourier transform (FT) moduli χ(R) as a function of radial distribution distance (R-φ) and the corresponding best fits for Cr$_{0.51}$N$_{0.49}$ thin film sample (a). FT moduli in the (R-φ) space for the Cr$_{0.51}$N$_{0.49}$, Cr$_{0.50}$V$_{0.03}$N$_{0.47}$ and Cr$_{0.44}$Mo$_{0.09}$V$_{0.08}$N$_{0.39}$ samples with different alloying concentrations (b).

Figure 2(a) shows the real part of the Fourier Transform (FT) moduli χ (R) and the corresponding best fit as a function of radial distribution distance (R-φ) for Cr$_{0.51}$N$_{0.49}$ thin film. The fitting of FT χ (R) of the Cr foil is shown and discussed in the supplementary information (Figure S1).[47] Single scattering theory with the first two nearest neighbor scattering paths from the Cr absorber atom were considered for fitting the FT χ (R) spectra. The first two shells extended in 0.7-3.1 Å (see Figure 2(a)) correlate to the Cr-N and Cr-Cr bond distances at around 1.55 and 2.45 Å. The scattering phase shift in EXAFS is typically 0.5 Å at lower (R-φ) from the obtained fitted values since χ (k) ∝ sin (kR+φ) in k-space.[52] The atomic pair distances (R$_{absorber-neighbor}$) obtained from the fitting are R$_{Cr-N}$ = 2.08 (±0.04) and R$_{Cr-Cr}$ = 2.92 (±0.01) Å. The R$_{Cr-N}$ value is in excellent agreement while R$_{Cr-Cr}$ is slightly at a lesser value compared to the XRD data (R'$_{Cr-N}$ = 2.079 and R'$_{Cr-Cr}$ = 2.941). This local information may differ from the macroscopically averaged information acquired from XRD.

The fitting reveal values of N$_{Cr-N}$ = 4.85 (±1.5) and N$_{Cr-Cr}$ = 10.6 (±1.1) in the first and second coordination shell, respectively. Contrary to our RBS results, from EXAFS, the N deficiency around the first coordination shell for Cr$_{0.51}$N$_{0.49}$ sample cannot be stated within the error bar of the present study. Fitting of the RBS data yields an overall substoichiometry in nitrogen if the elemental distributions are assumed to be homogeneous. Local inconsistencies from the stoichiometry can be probed using EXAFS which typically extends up to 5 Å from the Cr absorber atom. EXAFS fitting further suggests locally less occupancy of Cr atoms in the second coordination shell.

Figure 2(b) shows distinct variation in the intensity of oscillations of $Cr_{0.51}N_{0.49}$, $Cr_{0.50}V_{0.03}N_{0.47}$ and highest alloyed $Cr_{0.44}Mo_{0.09}V_{0.08}N_{0.39}$ samples around the first and second coordination shell (up to 3Å). The maximum intensity is observed at around 2.45 Å for $Cr_{0.51}N_{0.49}$ and it gradually decreases with increasing alloying concentrations. The peak area of the radial distribution distance is correlated to the coordination number.[53] Therefore, it can be inferred that for $Cr_{0.51}N_{0.49}$, the local coordination (N and next nearest Cr atoms) is the maximum for Cr absorber leading to highest intensity in both the shells.

For $Cr_{0.50}V_{0.03}N_{0.47}$ sample, the intensity of the first shell remains unaltered in comparison to $Cr_{0.51}N_{0.49}$. However, 3% alloying of V leads to substitution of few Cr atoms with V and is reflected in less Cr-Cr bonds seen from a reduced intensity around the second shell. For $Cr_{0.44}Mo_{0.09}V_{0.08}N_{0.39}$ sample, presence of less Cr-N and Cr-Cr bonds are evident from the radial distribution spectra compared to rest of the samples. This is due to the bond formation of Cr absorber atom with V and Mo atoms which substituted few Cr atoms in the metal site. Consequently, due to substoichiometry of N as observed from our RBS results, the resultant Cr-N bonds also reduce. Locally, the less directional Cr-Cr bond length reduces for $Cr_{0.44}Mo_{0.09}V_{0.08}N_{0.39}$ compared to rest of the samples demonstrating similar trend as our XRD results. This is attributed to the presence of N vacancies due to transition from covalent to metal-like character.[50]

### 3.3 Electronic structure of unoccupied and occupied states

*3.3.1 X-ray Absorption Near Edge Spectroscopy*

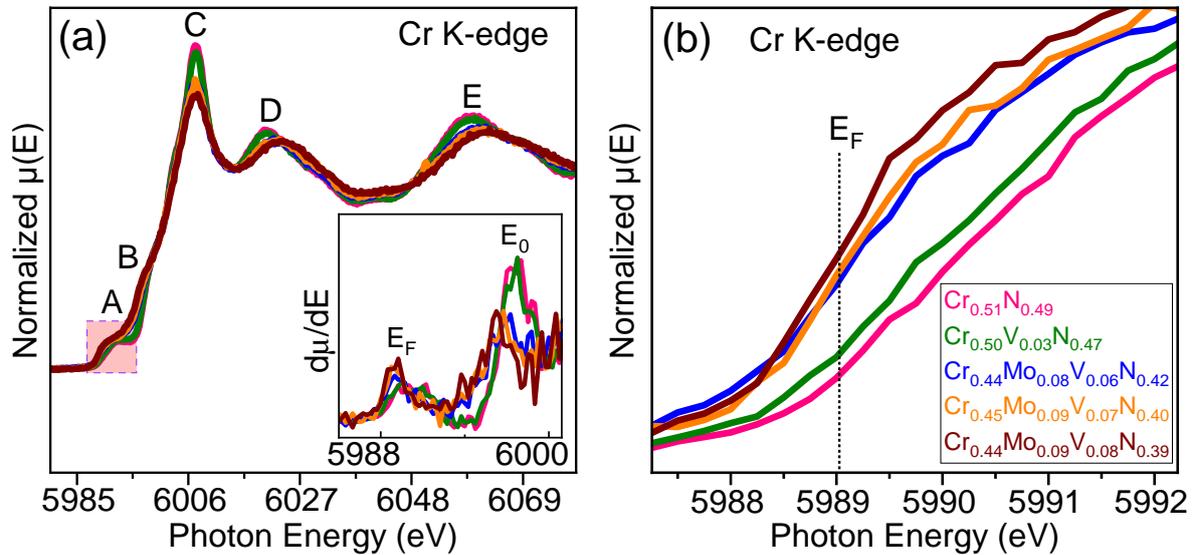

**Figure 3:** Normalized Cr *K*-edge XANES spectra of $Cr_{0.51}N_{0.49}$, $Cr_{0.50}V_{0.03}N_{0.47}$ and $Cr(Mo,V)N_x$ thin film samples in different alloying concentrations. The inset shows the first order derivative of the absorption spectra with respect to the photon energy. $E_F$ represents the Fermi level, and $E_0$ represents the absorption edge of the samples (a). The magnified view of the same normalized Cr K-edge XANES spectra near the pre-edge region indicating $E_F$ is shown for $Cr_{0.51}N_{0.49}$, $Cr_{0.50}V_{0.03}N_{0.47}$ and $Cr(Mo,V)N_x$ samples (b).

Figure 3(a) shows normalized Cr *K*-edge XANES spectra of $Cr_{0.51}N_{0.49}$, $Cr_{0.50}V_{0.03}N_{0.47}$ and $Cr(Mo,V)N_x$ samples. The observed features are labelled as A-E. For comparison, a reference Cr foil was also measured and shown in the supplementary information (Figure S2).[47] Note that all the samples display a weak pre-edge feature, shown in Figure 3(a) as a shaded region labeled

A. This is assigned to a single core electron excitation from the *1s* core orbital to the unoccupied *3d* valence states of the Cr absorber atom partially hybridized with the *2p* valence states of the neighboring N atoms and is electric dipole allowed ($\Delta l = \pm 1$). The $E_F$ lies around the pre-edge, which is indicated in the inset of Figure 3(a) (also in the magnified view of Figure 3(b)).

Above the Fermi energy, $E_F$, the higher-energy feature labeled B is attributed to *1s* → *4s* transitions of the Cr absorber atom, with partial contribution from *2p-3s-3p* states of the nitrogen ligand. This makes the electric dipole transition allowed ($\Delta l = \pm 1$) as observed for different transition metal compounds.[54] Around this region, the absorption edge ($E_0$) also appears (indicated in the inset of Figure 3(a)) and the positions for all the samples are listed in Table I. The $E_0$ gradually shifts to the lower photon energy from $Cr_{0.51}N_{0.49}$ to $Cr_{0.44}Mo_{0.09}V_{0.08}N_{0.39}$. This is attributed to the reduced core-hole screening of the Cr ions from $Cr_{0.51}N_{0.49}$ to $Cr_{0.44}Mo_{0.09}V_{0.08}N_{0.39}$ leading to reduction in the charge transfer from Cr to N. The trend in the samples can be well corroborated to the different electronegativities of the transition elements present in the samples with 1.63 (V), 1.66 (Cr), 2.16 (Mo), and 3.04 (N) in Pauling scale.

As seen from XRD, the samples crystallize in cubic rocksalt NaCl structure ($Fm\bar{3}m$). This structure is known to have a periodic ABCABC stacking sequence. Such a stacking sequence in the presence of a nitrogen ligand environment gives rise to an intense *white line* (a sharp intense peak in the near edge) feature labeled C. Hence, such observation is a fingerprint for the typical characteristics of transition metal nitrides crystallized in NaCl structure. Features labeled C and D arises due to core electron transition from *1s* → *4p* obeying the electric dipole transition rule. The occupancy of electrons in the unoccupied *4p* orbitals of the Cr absorber reflects an inverse trend in the intensity around feature C. Feature E is resultant of the constructive interference of the outgoing photoelectron from the Cr absorber and backscattered photoelectron wave function from the neighboring N atoms.

A magnified view of the electronic DOS around $E_F$ of the normalized Cr *K*-edge XANES spectra is shown in Figure 3(b). In contrast to stoichiometric CrN,[55] non-vanishing Cr *3d* $t_{2g}$ non-bonding and Cr *3d* $e_g$ anti-bonding states with partial contribution from N *2p* states arise around the $E_F$. This is due to the N vacancy mediated defects for the substoichiometric $Cr_{0.51}N_{0.49}$ in the present study and is consistent with previous band structure calculations.[29] The vacancy in the N site led to occupancy of electrons back to the metallic spin up Cr *3d* $t_{2g}$ non-bonding states pushing the $E_F$ inside the conduction band. The increase in the DOS around $E_F$ with alloying (also indicated around $E_F$ in the inset of Figure 3(a)) is governed by the cumulative effect of alloying elements and N vacancies. The effect is strongly correlated to the change in the thermoelectric properties of the samples (discussed in Section 3.4 below).

*3.3.2 Soft X-ray Absorption Spectroscopy*

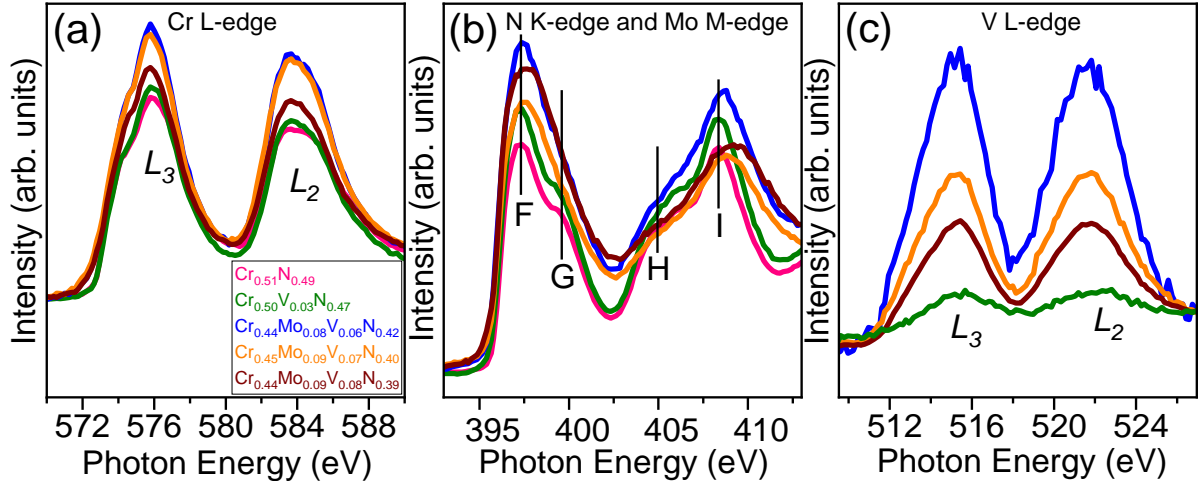

**Figure 4:** Normalized N *K*-edge and Mo $M_{3,2}$-edge (a), Cr $L_{3,2}$-edge (b) and V $L_{3,2}$-edge (c) XANES spectra of $Cr_{0.51}N_{0.49}$, $Cr_{0.50}V_{0.03}N_{0.47}$ and $Cr(Mo,V)N_x$ thin film samples with different alloying concentrations.

Figure 4 shows normalized XANES spectra of Cr $L_{3,2}$-edge, N *K*-edge and Mo $M_{3,2}$-edge, and V $L_{3,2}$-edge of $Cr_{0.51}N_{0.49}$, $Cr_{0.50}V_{0.03}N_{0.47}$ and $Cr(Mo,V)N_x$ thin film samples. For $Cr_{0.51}N_{0.49}$, the doublet features ($L_3$ and $L_2$) in the Cr $L_{3,2}$ spectrum is the resultant of spin orbit splitting ($\zeta_{sp}$) due to electric dipole allowed transition ($\Delta l = \pm 1$) of a core electron from Cr $2p_{3/2}$, $2p_{1/2}$ →Cr *3d* states owing to $\zeta_{sp} \sim 8.1$ eV. The lower energy shift of ~1 eV of $L_{3,2}$ peaks from previous report suggest $Cr^{+3-\delta}$ valence state in our sample. Although the line shape and $\zeta_{sp}$ value matches closely with stoichiometric CrN.[56] Upon alloying, the relative changes in the $E_0$ and $\zeta_{sp}$ values of the samples compared to $Cr_{0.51}N_{0.49}$ fall within the detection limit. Higher intensity around $L_3$ and $L_2$ for $Cr_{0.50}V_{0.03}N_{0.47}$ and $Cr(Mo,V)N_x$ samples compared to $Cr_{0.51}N_{0.49}$ can be partly explained by the presence of fewer *d*-electrons per atom of Cr upon Mo and/or V substitution leading to higher unoccupied Cr *3d* states.

For $Cr_{0.51}N_{0.49}$ and $Cr_{0.50}V_{0.03}N_{0.47}$ samples, four distinct features (F, G, H and I) in the N *K*-edge correspond to core electron transitions from the N *1s* core level to unoccupied hybridized states of: (F) N $2p\pi$+Cr *3d* i.e., $t_{2g}$; (G) N $2p\sigma$+Cr *3d* i.e., $e_g$; and (H and I) higher unoccupied hybridized states of N *2p*+Cr *4sp*. The features indicate strong hybridization between Cr *3d* and N *2p* states, yielding a crystal field splitting (10Dq) of ~2.1 eV. Usually this parameter has significance as it is anti-correlated to the lattice parameter [*i.e.,* 10Dq ∝ (bond distance)$^{-5}$] and indicates hybridization strength.[57] However, no significant changes in the 10Dq values between $Cr_{0.51}N_{0.49}$ and $Cr_{0.50}V_{0.03}N_{0.47}$ are observed in the present study within the energy resolution limit. For the $Cr(Mo,V)N_x$ system, an overlap between the N *K*-edge (401.6 eV) and the Mo $M_{3,2}$-edges (spread over 392 - 410 eV),[58] makes it complicated to analyze the N *K*-edge after addition of Mo. Due to the low fluorescence yield at the Mo $M_{3,2}$-edge,[59] the N *K*-edge dominates the XANES spectra. The noticeable suppression of features G and H are caused by the band smearing of unoccupied Mo *4d* states due to the opening of the Mo *3p*→*4d* dipole allowed transition channel. The broadening also overshadows the relative chemical shifts of $E_0$ between the samples.

Similar to the Cr *L*-edge spectra, the same trend can be observed in the V *L*-edge spectra, where the $L_3$ and $L_2$ features start appearing for $Cr_{0.50}V_{0.03}N_{0.47}$ and becomes prominent for the $Cr(Mo,V)N_x$ samples. However, in a simple electronic picture with higher V alloying, V should contribute with more *3d* states. Simultaneously more electrons should be drawn from V to N because of the lower electronegativity, provided the N content is constant. Although at first glance it may appear simple, the presence of three transition metals with a gradual reduction of N/Me ratio is a complex system. A smaller number of available N *2p* states enforces a competition of the charge transfer from one of the metals to N and back to the other metal site, especially within the Cr(Mo,V)N$_x$ series, depending on the electronegativities of the metals. This affects the intensity distribution, and therefore no definite trend within the Cr(Mo,V)N$_x$ system can be observed from all the absorption edges.

*3.3.3 X-ray Photoelectron Spectroscopy*

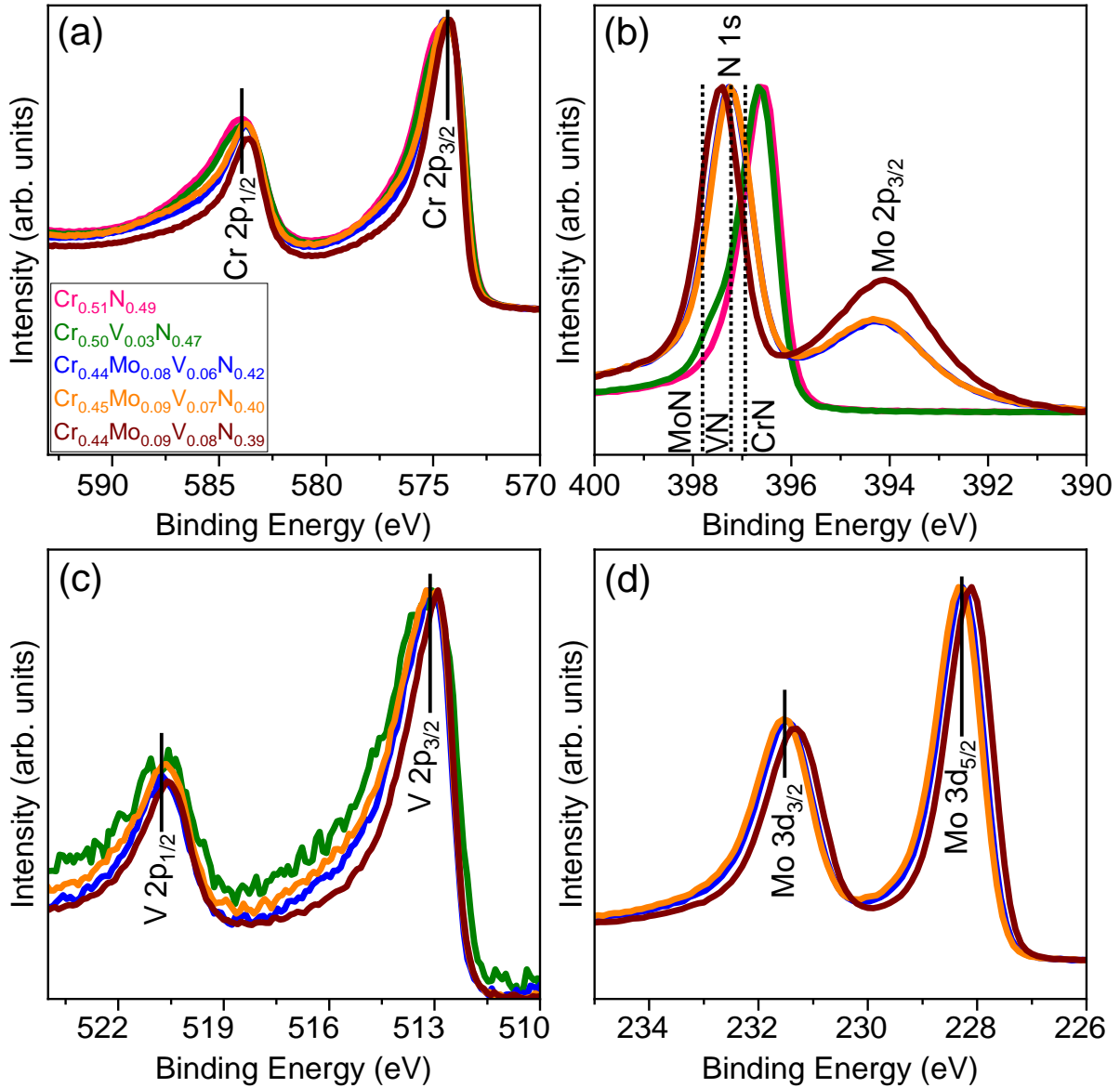

**Figure 5:** XPS core level spectra of Cr *2p* (a), N *1s* and Mo *2p* (b), V *2p* (c), and Mo *3d* (d) of Cr$_{0.51}$N$_{0.49}$, Cr$_{0.50}$V$_{0.03}$N$_{0.47}$ and Cr(Mo,V)N$_x$ samples with different alloying concentrations.

Figure 5 shows the Cr *2p*, N *1s*, Mo *2p*, V *2p*, and Mo *3d* XPS core level spectra normalized to the highest intensity. Cr *2p* core level spectra (see Figure 5(a)) reveal the spin-split doublet peaks *2p$_{3/2}$* and *2p$_{1/2}$*. For Cr$_{0.51}$N$_{0.49}$ and Cr$_{0.50}$V$_{0.03}$N$_{0.47}$, the peaks *2p$_{3/2}$* and *2p$_{1/2}$* are centered at around 574.4 and 583.9 eV. An asymmetry around the main *2p$_{3/2}$* peaks is noted which arises in the Cr photoelectron spectrum owing to the multiplet structure due to unpaired electrons.[60] Around the broadened *2p$_{1/2}$* peaks, such effect is less pronounced due to the *Coster-Kronig* effect.[61] For Cr(Mo,V)N$_x$ samples, no notable peak shifts can be observed compared to Cr$_{0.51}$N$_{0.49}$ and Cr$_{0.50}$V$_{0.03}$N$_{0.47}$. This indicates no significant change in the valence charge distribution of Cr ions after alloying which is contrary to the observed absorption edge shift in Cr K-edge XANES spectra. Thus, the deviation can be understood in the difference of the probed volume in both the measurements. However, reduction in peak broadening and peak asymmetry can be observed with Mo and increasing V alloying concentrations.

In Figure 5(b), the N *1s* and Mo *2p$_{3/2}$* partially overlapping peaks are shown. The dotted lines show the reference position of the N *1s* spectra for CrN, VN and MoN, respectively.[62] For all samples, including even Cr$_{0.51}$N$_{0.49}$, the N *1s* peak shifts to lower BE from the reference value reported for stoichiometric CrN (396.9 eV). This is due to the N substoichiometry (N/metal ratio is 0.96), which results in that each N atom has on average more Cr neighbors. That can lead to both (i) higher negative charge density on each Cr atom, and (ii) better screening of the core hole left after photoemission. Both effects result in the peak shift to lower BE. A gradual shift of N *1s* core level spectra to the higher BE side can be seen for Cr$_{0.50}$V$_{0.03}$N$_{0.47}$ and Cr(Mo,V)N$_x$ samples. The observation is in line with the same trend of shift from reference samples. Thus, the shift is due to gradual reduction of N content in the non-metal site and addition of other transition metals *i.e.*, Mo and V in the CrN matrix. The result implies reduction in the charge state of N. Although it should be noted that in XPS there is a probability of preferential sputtering of N during sputter cleaning of the sample surface.

V *2p* core-level spectra shown in Figure 5(c) reveal no significant peak shifts. The only visible change is the reduction in peak asymmetry that takes place with increasing V content. Mo *3d* core-level spectra (see Figure 5(d)) from Cr$_{0.44}$Mo$_{0.08}$V$_{0.06}$N$_{0.42}$ and Cr$_{0.45}$Mo$_{0.09}$V$_{0.07}$N$_{0.40}$ films are identical. A shift to the lower BE side (~0.2 eV) can be noted in the Mo 3d spectrum from the Cr$_{0.44}$Mo$_{0.09}$V$_{0.08}$N$_{0.39}$ sample. As the corresponding N 1s peak shifts to higher BE (see Figure 5(b)) this corroborates a reduced charge transfer from metal to N atoms.

The $\zeta_{sp}$ value of 9.5 and 7.7 eV is obtained for Cr *2p* and V *2p* XPS core level spectra. A discrepancy of 1.4 eV for Cr and 1.2 eV for V, among the $\zeta_{sp}$ values between the XAS and XPS measurements can be noted. This is due to excitation of the electrons to different final states involved in both the processes. Considering *2p$^6$3d$^n$* as the ground state, the final states for XAS and XPS are *2p$^5$3d$^{n+1}$* and *2p$^6$3d$^n$*, respectively. It leads to variable exchange and Coulomb interaction between the transition states involved resulting in such discrepancy.[63] theoretical *total*-density of states calculations (Figure 6 (b), (c) and (d)) reveal contribution from the different states in the valence band spectra.

*3.3.4 Valence Band Spectroscopy*

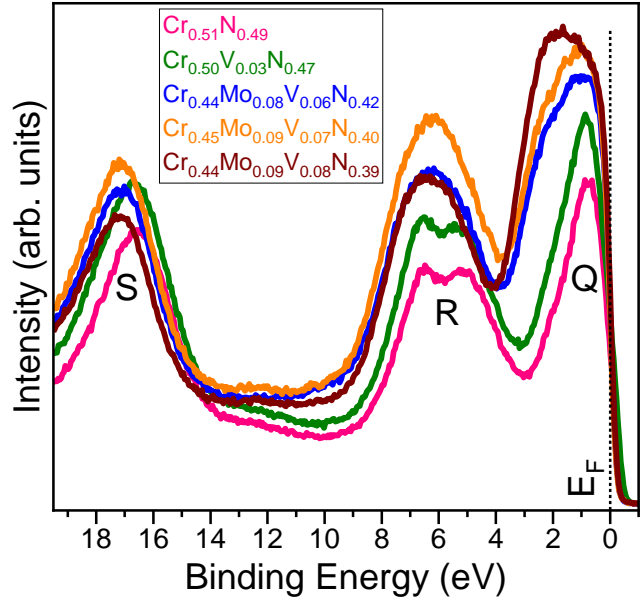

**Figure 6:** Valence band spectra of $Cr_{0.51}N_{0.49}$, $Cr_{0.50}V_{0.03}N_{0.47}$ and $Cr(Mo,V)N_x$ samples with different alloying concentrations.

Figure 6 shows valence band spectra for the $Cr_{0.51}N_{0.49}$, $Cr_{0.50}V_{0.03}N_{0.47}$ and the $Cr(Mo,V)N_x$ samples with their indicated $E_F$. In the supplementary information, the theoretical *total*-DOS calculations are shown in Figure S3.[47] Note that double layer antiferromagnetic ordering with a Hubbard parameter $U=3$ was considered in the calculations.[64] The calculations were used to identify the hybridization contributions and their positions in the valence band spectra.

In contrast to stoichiometric CrN known to exhibit a narrow band gap,[55,65] finite DOS around $E_F$ in the present study is attributed to the presence of N vacancies in $Cr_{0.51}N_{0.49}$ thin film sample. From previous studies it is known theoretically that N vacancies induce *n*-type behavior in the DOS.[29] However, in this study the x-ray width is ~0.3 eV. Feature Q arises mainly due to the contribution from the Cr *3d* states hybridized with N *2p* states, whereas feature R around 5 - 8 eV is an effect from Cr *3d* states hybridized to N *2p* states with partial contribution from Cr *3p* states. The shoulder (at 6.5 eV) arises as intense as the main feature J (5.2 eV) and appears as a doublet.[66] The contribution of feature S is mostly dominated by N *2s* states with a small contribution from Cr *3d* states.

For $Cr_{0.50}V_{0.03}N_{0.47}$ sample, the features of the valence band spectra reciprocate similar trend like $Cr_{0.51}N_{0.49}$. However, calculated *total*-DOS reveals essential contribution from both Cr *3d* and N *2p* states with partial contribution from V *3d* states for feature Q in this sample.[32] Feature R unveils hybridization of N *2p* states with (Cr,V) *3d* and small contribution from Cr *3p* states. Feature S is contributed due to the hybridization of N *2s* states with marginal contribution from Cr *3d* states.

For $Cr(Mo,V)N_x$ samples, the main contribution of the features is due to the hybridization of the following states:

- Feature Q at around 0.8 - 1.6 eV: N *2p* states - Cr *3d*, V *3d* and primarily Mo *4d* states.
- Feature R around 5 - 8 eV: N *2p* states - Cr (*3d,3p*)/V (*3d,3p*)/ Mo *4d* states.
- Feature S around 16 eV: N *2s* states – small contribution from Cr (*3p,3d*)/Mo *4d* states.

Here, a band smearing across feature Q is due to overlap of the broadened *4d* wavefunctions of Mo with (Cr,V) *3d* wavefunctions leading to delocalization of the valence band. The effect is also pronounced around feature R with diminished doublet feature as was observed for $Cr_{0.51}N_{0.49}$ and $Cr_{0.50}V_{0.03}N_{0.47}$. For feature S, a shift to lower BE (0.5 eV) can be observed for $Cr(Mo,V)N_x$ samples compared to $Cr_{0.51}N_{0.49}$ and $Cr_{0.50}V_{0.03}N_{0.47}$.

*3.3.5 Resonant Inelastic X-ray Scattering*

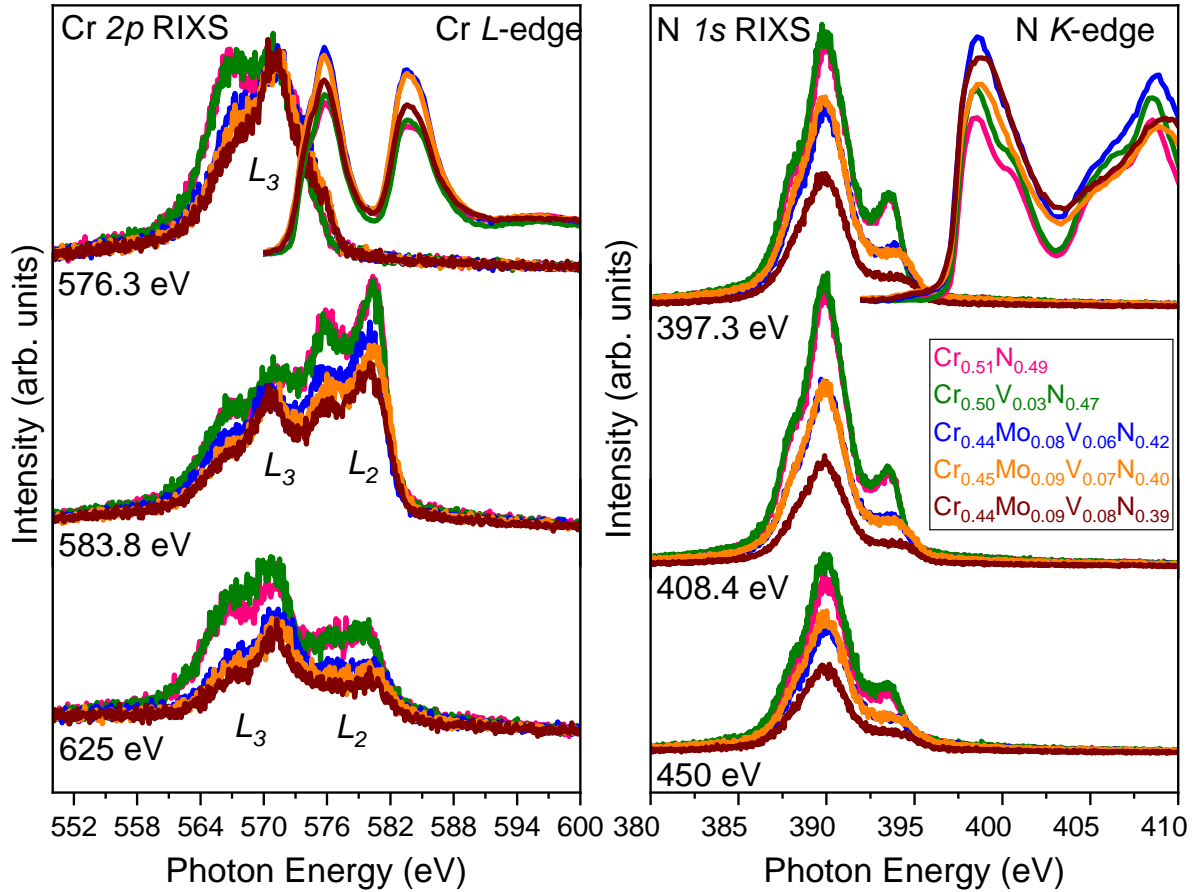

**Figure 7:** Cr *2p* RIXS at different excitation photon energies of 576.3 (resonant), 583.8 (resonant) and 625 eV (non-resonant), respectively, and Cr $L_{3,2}$-edge XAS data (top right) (a), N *1s* RIXS at different excitation photon energies of 397.3 (resonant), 408.4 (resonant) and 450 eV (non-resonant), respectively and N *K*-edge XAS (b) of the $Cr_{0.51}N_{0.49}$, $Cr_{0.50}V_{0.03}N_{0.47}$ and the $Cr(Mo,V)N_x$ thin film samples.

Figure 7(a) shows Cr *2p* RIXS, Cr $L_{3,2}$-edge XAS data and Figure 7(b) shows N *1s* RIXS, N *K*-edge XAS data, respectively of the $Cr_{0.51}N_{0.49}$, $Cr_{0.50}V_{0.03}N_{0.47}$ and the $Cr(Mo,V)N_x$ thin film samples. The peak maxima in the XAS data were used to determine the photon energies for the emission measurements. Cr *2p* RIXS on a Cr thin metal film is shown in the supplementary information (Figure S4) to obtain insight on the *partial*-DOS and a comparison between Cr and $Cr_{0.51}N_{0.49}$.[47] The Cr *2p* RIXS spectra represents the *sd*-DOS of the occupied Cr states of the valence band following the Cr *3d4s* → *2p$_{3/2,1/2}$* dipole transitions ($\Delta l = \pm 1$).

The spectra excited at 576.3 eV (resonant) show doublet features in the valence band region. In contrast, the spectra excited at 583.8 eV (resonant) and 625 eV (non-resonant) exhibit four features. For all samples, we attribute these features in the valence band region as $L_3$ and $L_2$ emission with a $t_{2g}$-$e_g$ sub splitting. The most intense $L_3$ emission at 576.3 eV and $L_2$ emission at 583.8 eV are due to excitation at the *2p₃/₂* and *2p₁/₂* absorption edges. The feature at the lowest emission energy arise primarily due to *3d t₂g* orbitals, with partial N *2p* contribution, whereas at higher emission energy, the *3d eg* orbitals with admixture of N *2p* orbitals dominates.[64,67]

At 576.3 eV excitation energy, the relatively large distance of the band maxima from the crossover of the RIXS and XAS spectra of this bond region is an indication of strong covalent bonding between Cr and N. However, the doublet feature is less pronounced for the Cr(Mo,V)N$_x$ samples with reduced bandwidth. It indicates a decrease in the Cr *3d* - N *2p* hybridization with reduced states leading to less covalent bonds. For 583.8 excitation energy, a reduced intensity of the Cr(Mo,V)N$_x$ samples is observed compared to the Cr$_{0.51}$N$_{0.49}$, and the Cr$_{0.50}$V$_{0.03}$N$_{0.47}$ samples, in-line with the reduced atomic % of Cr from our RBS measurements (see Table I). At non-resonant 625 eV excitation energy, the *L₃/L₂* branching ratio increases from 1.9 for Cr$_{0.51}$N$_{0.49}$ to 4.4 for Cr$_{0.44}$Mo$_{0.09}$V$_{0.08}$N$_{0.39}$. Quantitatively, the significantly higher *L₃/L₂* ratio for the Cr$_{0.44}$Mo$_{0.09}$V$_{0.08}$N$_{0.39}$ sample compared to the statistical ratio (2:1) is due to the more effective Coster-Kronig process in conducting systems compared to more localized electrons with less conduction of Cr$_{0.51}$N$_{0.49}$ sample.[68]

The N *1s* RIXS spectra representing the *partial*-DOS of N valence region follows the *2p* → *1s* dipole transitions. The N *1s* RIXS spectra excited of all the samples at resonant photon energy of 397.3 eV exhibits a main peak centered around 389.9 eV composed of primarily N *2p* states, in agreement with band structure calculations.[69] We interpret the low-energy emission shoulder at ~388 eV below the main peak as N *2s-2p* hybridization. A higher-energy shoulder at ~393.6 eV is also observed attributed to N *2p* states hybridized with Cr *3d* states in line with theoretical density functional theory calculations.[55] The intensity of this shoulder is highest for the Cr$_{0.51}$N$_{0.49}$ and the Cr$_{0.50}$V$_{0.03}$N$_{0.47}$ samples due to more directional bonds and more charge withdrawal from Cr and V to N when there is no Mo content. For the Cr(Mo,V)N$_x$ samples containing Mo, the number of hybridized N *2p* states around the crossover region are significantly higher. This in turn affects the electrical resistivity within the samples as discussed in the next section. Contrary to the Cr *2p* RIXS, most excitation-energy dependent changes in the N *1s* RIXS are only observed in the high-energy shoulder, while there are only minor changes in the main peak. This is a signature of delocalized N *2p* states compared to the more localized Cr *3d* states.

3.4 Seebeck Coefficient and Electrical Resistivity

**Table II**. The Seebeck coefficient, electrical resistivity, and power factor of the samples.

| Samples | S (µV K$^{-1}$) | ρ (µΩcm) | S²σ ± 25% (µW cm$^{-1}$ K$^{-2}$) |
|---|---|---|---|
| Cr$_{0.51}$N$_{0.49}$ | -93 ± 1.0 | 26722 ± 2820 | 0.3 |
| Cr$_{0.50}$V$_{0.03}$N$_{0.47}$ | -95 ± 0.9 | 4460 ± 1000 | 2 |
| Cr$_{0.44}$Mo$_{0.08}$V$_{0.06}$N$_{0.42}$ | -14 ± 0.6 | 314 ± 3 | 0.6 |
| Cr$_{0.45}$Mo$_{0.09}$V$_{0.07}$N$_{0.40}$ | -9 ± 0.3 | 253 ± 2 | 0.3 |
| Cr$_{0.44}$Mo$_{0.09}$V$_{0.08}$N$_{0.39}$ | -7 ± 0.4 | 192 ± 5 | 0.2 |

The Seebeck coefficient and electrical resistivity (ρ) of Cr$_{0.51}$N$_{0.49}$, Cr$_{0.50}$V$_{0.03}$N$_{0.47}$ and Cr(Mo,V)N$_x$ samples are tabulated in Table II. For Seebeck coefficient, equation (1) can be re-written as,[70]

$$S = \frac{8\pi^2 k_B^2 T}{3qh^2} m^* \left(\frac{\pi}{3n}\right)^{2/3} \ldots\ldots\ldots\ldots(4)$$

Combining equation (4) and (2), Seebeck coefficient and electrical conductivity are interrelated. For $Cr_{0.51}N_{0.49}$, presence of small amount of N vacancies results in S value of -93 µV K$^{-1}$ at room temperature which is at least 3 times higher than earlier reports without any post deposition treatment.[17,49] Rather ρ and thermoelectric power factor $S^2\sigma$ value also seems to be at per with stoichiometric bulk CrN.[71] We attribute the enhanced performance is due to the presence of sharp and local increase in the DOS near $E_F$ in the VBS spectra (see Figure 6) in line with the estimated theoretical band structure calculations.[21] For $Cr_{0.50}V_{0.03}N_{0.47}$, S remains nearly same but the abrupt decrease in the ρ value is correlated to the increased population across the $E_F$, as seen from our XANES and VBS study (See Figure 3(b) and 6) resulting 6 times higher value in power factor compared to $Cr_{0.51}N_{0.49}$. However, for $Cr(Mo,V)N_x$ series, a large reduction in electrical resistivity with typical metal like S values originates from the strong hybridization of the N *2p*-(Cr,V) *3d* states with Mo *4d* states inducing higher DOS across the $E_F$ as evidenced from our RIXS, XAS and VBS studies (See Figure 7(a)). Moreover, a strong coupling between Mo *4d* and N *2p* states near the Fermi level weakens the Cr *3d*-N *2p* electronic correlations driving it far from Mott insulator.[33] However, the power factor of $Cr(Mo,V)N_x$ series is still comparable to $Cr_{0.51}N_{0.49}$ which we attribute to the N substoichiometry.

## Conclusions

In summary, we systematically studied the effect of V and/or Mo alloying in the CrN matrix, with substoichiometric N as revealed by our RBS study. For the first time we successfully synthesized epitaxial single phase $Cr(Mo,V)N_x$ multicomponent nitride. The addition of V stabilizes the cubic phase retention in this complex system despite the presence of higher atomic % of Mo. Even minuscule N substoichiometry led to diminished band gap in $Cr_{0.51}N_{0.49}$ due to lesser N 2p states available to accommodate the electrons and instead return to the metal site shifting the Fermi level towards the conduction band. For $Cr_{0.50}V_{0.03}N_{0.47}$, less N content and marginal alloying of V leads to lower hybridization of the Cr *3d*-N *2p* states revealing lower electrical resistivity without altering the Seebeck coefficient. This results in overall improvement of the thermoelectric power factor. Hence, it can be inferred that presence of N deficiency up to a critical limit still retains good thermoelectric properties. Later, in $Cr(Mo,V)N_x$ series, combined effect of N substoichiometry and contribution of Mo *4d* hybridized to N *2p* states weakens the Cr *3d*-N *2p* electronic correlations driving it far from Mott insulator. It is governed by crossover of significant density of states across the Fermi level compared to $Cr_{0.51}N_{0.49}$ and $Cr_{0.50}V_{0.03}N_{0.47}$ exhibiting metal-like resistivity. The N substoichiometry also leads to a reduction in the charge transfer from metal to N site. The reduced Seebeck coefficient of $Cr(Mo,V)N_x$ stems from presence of broadened Mo *4d* wavefunctions which drives it away from sharp and local increase in the density of states just below the Fermi level. Thus the present study shows potential of $Cr(Mo,V)N_x$ as a thermoelectric material which are strongly correlated to the density of states present near the Fermi level. The study motivates further research on N-stoichiometric Cr(Mo,V)N with lower alloying concentration of Mo for enhancement of the thermoelectric properties.

## Acknowledgements

The authors acknowledge funding from the Swedish Government Strategic Research Area in Materials Science on Functional Materials at Linköping University (Faculty Grant SFO-Mat-LiU No. 2009 00971), the Knut and Alice Wallenberg foundation through the Wallenberg Academy Fellows program (KAW-2020.0196), the Swedish Research Council (VR) under Project No. 2021-03826. M.M. also acknowledges financial support from the Swedish Energy Agency (Grant No. 43606-1) and the Carl Tryggers Foundation (CTS20:272, CTS16:303, CTS14:310). RS acknowledges support from the Swedish Research Council VR International Postdoc Grant 2022-00213 and the IUVSTA through the Medard W. Welch International Scholarship 2022. GG acknowledges the Swedish Energy Agency project 51201-1, the Åforsk Foundation Grant 22-4, and the Olle Enqvist foundation grant 222-0053.

Research conducted at MAX IV, a Swedish national user facility, is supported by the Swedish Research council under contract 2018-07152, the Swedish Governmental Agency for Innovation Systems under contract 2018-04969, and Formas under contract 2019-02496. The FEFF calculations were enabled by resources provided by the National Academic Infrastructure for Supercomputing in Sweden (NAISS) at Linköping University funded by the Swedish Research Council through grant agreement no. 2022-06725. Daniel Primetzhofer from Uppsala University is acknowledged for Accelerator operation supported by Swedish Research Council VR-RFI (Contract No. 2019-00191) and the Swedish Foundation for Strategic Research (Contract No. RIF14-0053).

# Thermoelectric properties and electronic structure of Cr(Mo,V)N$_x$ thin films studied by synchrotron and lab-based X-ray spectroscopy

Susmita Chowdhury*, Victor Hjort, Rui Shu, Grzegorz Greczynski, Arnaud le Febvrier, Per Eklund, and Martin Magnuson

*Thin Film Physics Division, Department of Physics, Chemistry and Biology (IFM), Linköping University, Linköping SE-581 83, Sweden*

## Supplementary Information

*1. X-Ray reflectivity (XRR)*

The XRR measurements were performed using PANalytical X'Pert diffractometer in line mode. For incident optics, a hybrid-mirror module with 0.5° divergence slit whereas in the diffraction optics side a 0.125° divergence slit was used. The PANalytical X'Pert Reflectivity software was used to fit the XRR data, and the obtained thickness and density of the samples are tabulated in Table S1. The thickness of reference samples Cr$_{0.51}$N$_{0.49}$ and Cr$_{0.5}$V$_{0.03}$N$_{0.47}$ are 45 and 56 nm, respectively while for Cr(Mo,V)N$_x$ samples the thickness are slightly at lower values of 27-33 nm. The obtained density of Cr$_{0.51}$N$_{0.49}$ is found to be less at 5.7 g cm$^{-3}$ compared to stoichiometric CrN (6.14 g cm$^{-3}$).[1] Due to gradual increase of alloying elements in the metal site and decrease in the N content, a trend of increase in the density can be observed for the alloyed samples.

**Table S1**. The thickness and density of the thin film samples obtained from fitting the XRR spectra.

| Samples | XRR analysis | |
|---|---|---|
| | Thickness (±2) (nm) | Density (g/cm³) |
| Cr$_{0.51}$N$_{0.49}$ | 45 ± 1 | 5.7 ± 0.1 |
| Cr$_{0.5}$V$_{0.03}$N$_{0.47}$ | 56 ± 2 | 6.2 ± 0.6 |
| Cr$_{0.44}$Mo$_{0.08}$V$_{0.06}$N$_{0.42}$ | 33 ± 1 | 6.4 ± 0.2 |
| Cr$_{0.45}$Mo$_{0.09}$V$_{0.07}$N$_{0.40}$ | 30 ± 1 | 6.5 ± 0.1 |
| Cr$_{0.44}$Mo$_{0.09}$V$_{0.08}$N$_{0.39}$ | 27 ± 1 | 6.4 ± 0.1 |

*2. Extended X-ray Absorption Fine Structure of Cr foil and Cr$_{0.51}$N$_{0.49}$ sample*

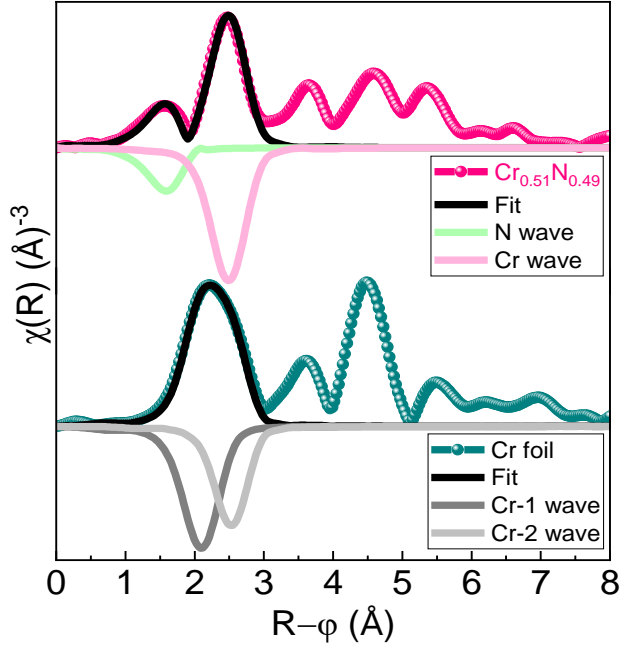

**Figure S1**. The Fourier transform moduli χ(R) as a function of radial distance (R-φ) and the corresponding best fits for metallic Cr foil and $Cr_{0.51}N_{0.49}$ thin film sample.

Figure S1 shows the real part of the Fourier Transform (FT) moduli χ (R) and the corresponding best fit as a function of radial distance (R-φ) for Cr foil and $Cr_{0.51}N_{0.49}$ thin film sample. As shown in Figure S1, the first two nearest neighbor Cr-Cr subshells constitute the first maxima in the FT spectra distributed over 1-3 Å in the (R-φ) range for the reference Cr foil. The corresponding best fits are tabulated in Table S2. For $Cr_{0.51}N_{0.49}$ sample, the first two shells extended in 0.7-3.1 Å correlate to the Cr-N and Cr-Cr bond distances at around 2.08 and 2.92 Å. A clear distinction between the FT spectra and the fitting parameters confirms the body centered cubic crystal structure of Cr foil and NaCl rocksalt structure of $Cr_{0.51}N_{0.49}$.

**Table S2**. The different structural metrical parameters obtained after fitting the FT spectra from 0-13.55 Å$^{-1}$ in k-region and 1-3 Å in the (R-φ) region. Here, $N_1$ and $N_2$ = first and second nearest neighbor co-ordination, $R_1$ and $R_2$ = atomic pair distance of the first and second neighbors i.e., Cr-N and Cr-Cr in case of CrN, $\sigma_1^2$ and $\sigma_2^2$= Debye Waller factor obtained from fitting of the first shell for Cr-foil and from first and second shell for $Cr_{0.51}N_{0.49}$ sample.

| Sample | $R_1$ | $N_1$ | $\sigma_1^2$ | $R_2$ | $N_2$ | $\sigma_2^2$ |
|---|---|---|---|---|---|---|
| Cr foil ($E_0$ = 5989.01) | 2.472 (±0.003) | 5.47 (±0.05) | 0.0072 (±0.000008) | 2.893 (±0.002) | 6 (0.02) | 0.0073 (±0.0007) |
| $Cr_{0.51}N_{0.49}$ ($E_0$ = 5997.5) | 2.08 (±0.04) | 4.85 (±1.5) | 0.006 (±0.005) | 2.92 (±0.01) | 10.6 (±1.1) | 0.0060 (±0.0008) |

*3. X-ray Absorption Near Edge Spectroscopy of Cr foil and $Cr_{0.51}N_{0.49}$ sample*

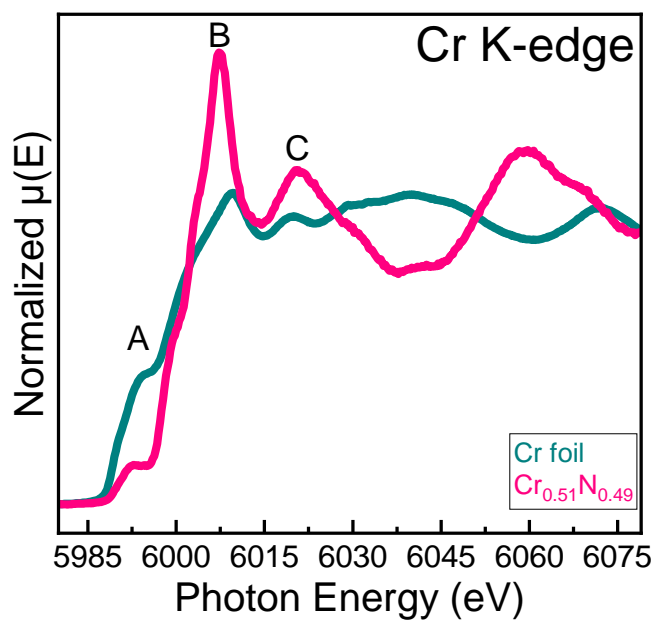

**Figure S2.** The normalized Cr K-edge XANES spectra of Cr foil and $Cr_{0.51}N_{0.49}$ thin film samples.

Figure S2 shows the XANES spectra of reference Cr foil and $Cr_{0.51}N_{0.49}$ thin film sample. In the present case, we approximated the absorption energy values typically around 50% of the edge jump, corresponding to the inflection point (feature A). The higher absorption energy shift of $Cr_{0.51}N_{0.49}$ compared to reference Cr foil indicates higher oxidation state of Cr in $Cr_{0.51}N_{0.49}$. The shift stems from the higher core hole screening of Cr ions in $Cr_{0.51}N_{0.49}$ due to charge transfer from Cr to N atoms as opposed to $Cr^0$ in Cr foil. The XANES spectra of Cr foil present different features compared to the nitride sample. This can be understood in terms of the different crystal structures owing to different local bonding environment of the nitride samples compared to the metallic Cr foil. Pure Cr exhibits body centered cubic crystal structure with a single stacking sequence, as it is not a closed packed structure. The $Cr_{0.51}N_{0.49}$ crystallize in cubic rocksalt NaCl-type structure ($Fm\bar{3}m$), with a periodic ABCABC stacking sequence. Such a stacking sequence in the presence of a nitrogen ligand environment gives rise to an intense *white line* feature B. It arises due to *1s → 4p* dipole transitions, compared to the less pronounced feature of metallic Cr foil. Here, the Cr foil has a lower intensity than $Cr_{0.51}N_{0.49}$. Charge transfer leads to higher empty unoccupied states in the *4p* orbitals of $Cr_{0.51}N_{0.49}$ than Cr metal which results in higher intensity around feature B and C.

*4. Theoretical density of states calculations*

(a)
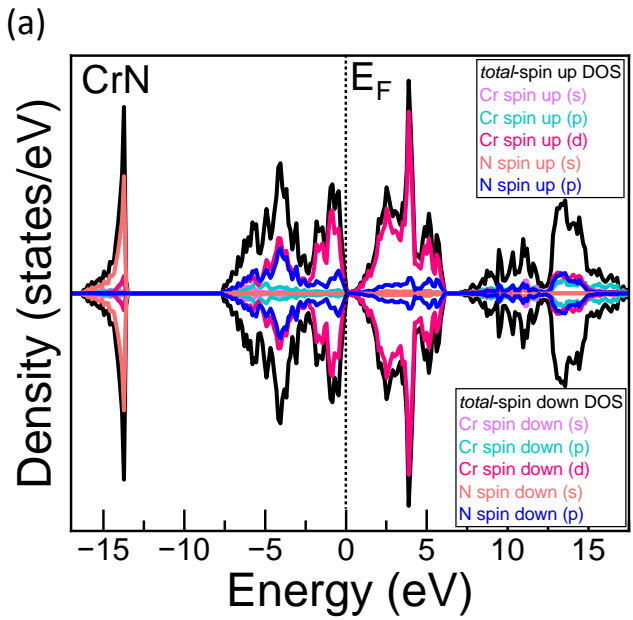
(b)
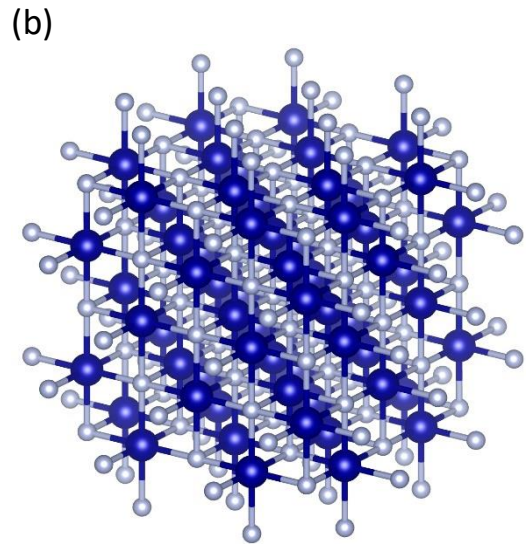

(c)
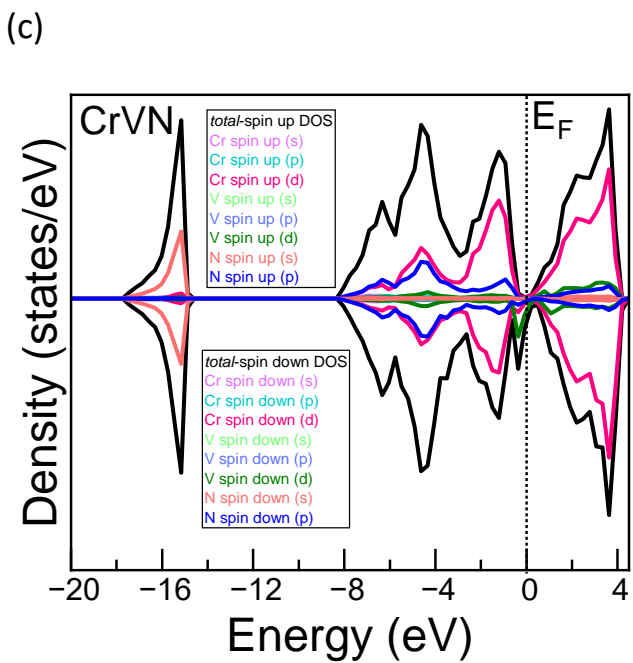
(d)
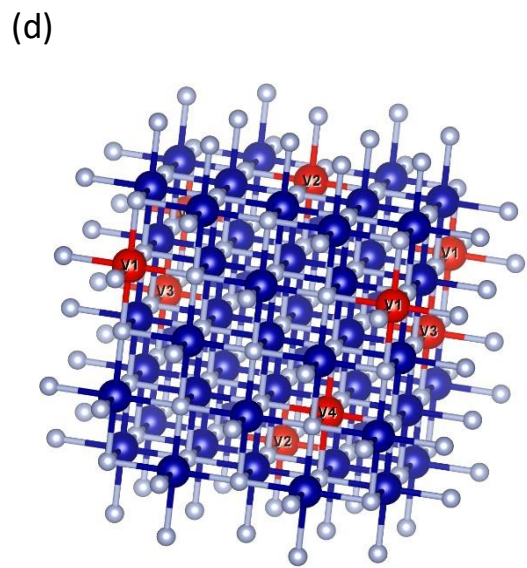

(e)
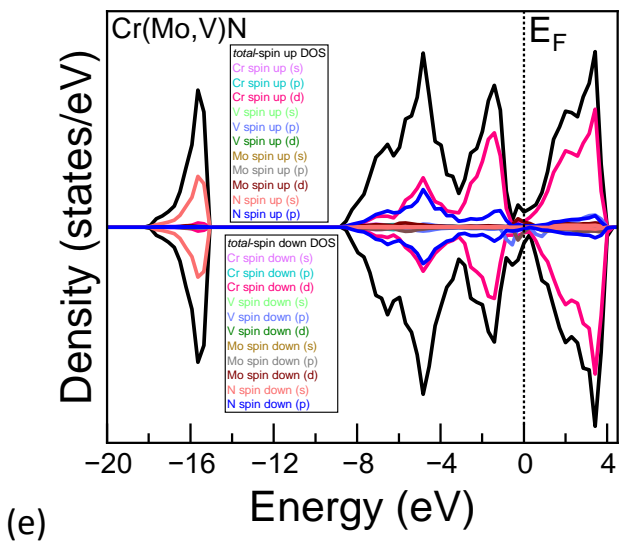
(f)
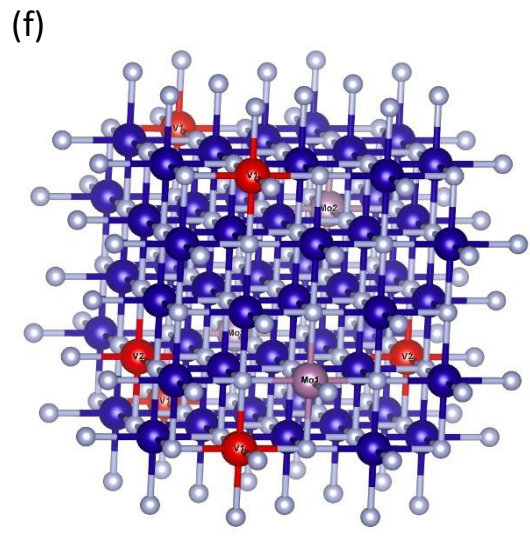

**Figure S3.** The *partial* and *total*-density of states calculated by *Density Functional Theory* and lattice structure of CrN (a, b), CrVN (c, d), and Cr(Mo,V)N (e, f) system.

Figure S3 shows the theoretical *partial* and *total* density of states (DOS) calculations (Figure S3 (a), (c) and (e)) revealing contribution from the different electronic states from -20 eV below to 15 eV above the Fermi level. Figure S3 (b), (d) and (f) shows the B1 NaCl rocksalt-type crystal structure of the theoretically calculated CrN, CrVN and CrVMoN. The electronic structure calculations were performed within a density-functional-theory framework and the projector augmented wave (PAW) method[2] as implemented in the Vienna ab initio simulation package(VASP)[3,4]. LDA[5] with a combination of a Hubbard Coulomb term (LDA+ U)[6,7] was used for treating electron exchange-correlation effects. The Hubbard terms U=3 eV[8] was applied to the Cr 3d, V 3d and Mo 4d orbitals. The energy cutoff for plane waves included in the expansion of wave functions was 400 eV. The CrN cell contained 32 Cr atoms and 32 N atoms, the CrVN cell contained 28 Cr atoms, 4 V atoms and 32 N atoms while the CrVMoN contained 28 Cr atoms, 2 V atoms, 2 Mo atoms and 32 N atoms. Sampling of the Brillouin zone was done using a Monkhorst-Pack scheme[9] on a grid of 5×5×5 (64-atom supercells) k points. The calculations were only used to identify the hybridization contributions in the valence band spectra. Note in the theoretical calculations stoichiometric crystal system are considered. However, our thin film samples exhibited N substoichiometry in $Cr_{0.51}N_{0.49}$, $Cr_{0.5}V_{0.03}N_{0.47}$ and $Cr(Mo,V)N_x$ samples. The details of the hybridized states have been compared to the experimentally obtained valence band spectra and discussed therein.

*5. Resonant Inelastic X-ray scattering of Cr foil and $Cr_{0.51}N_{0.49}$ thin film sample*

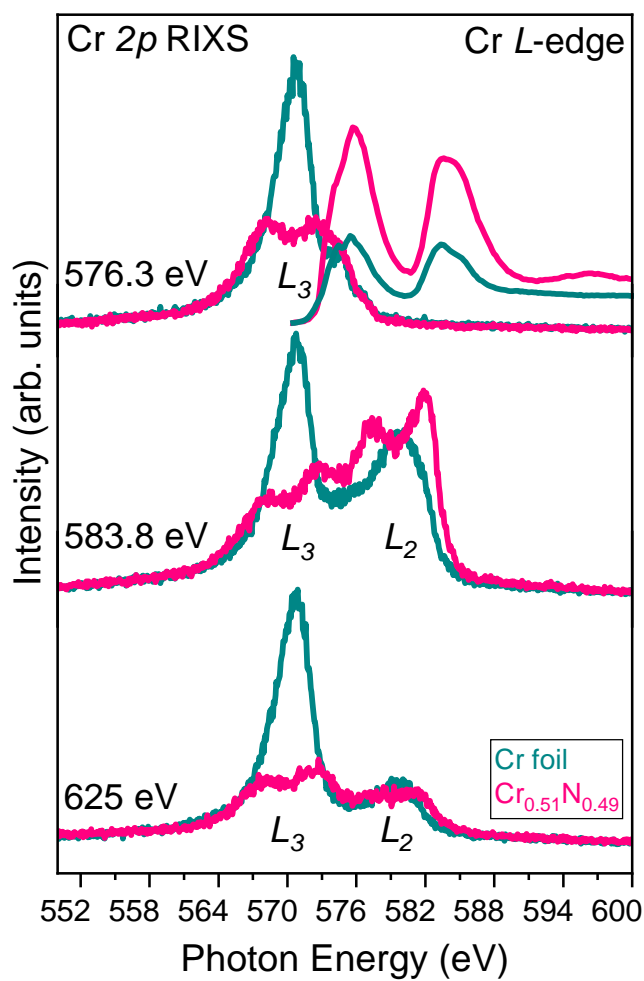

**Figure S4.** Resonant and non-resonant Cr *2p* RIXS data and Cr $L_{3,2}$-edge XAS data (top right) of the Cr foil and $Cr_{0.51}N_{0.49}$ thin film sample. All spectra are plotted on a photon energy scale.

Figure S4 shows Cr *2p* RIXS measured at 576.3, 583.8, and 625.0 eV photon energies, respectively and Cr $L_{3,2}$ XAS data of the Cr foil and $Cr_{0.51}N_{0.49}$ thin film sample. The Cr $L_{3,2}$ RIXS spectra follows the *3d4s* → *2p3/2,1/2* dipole transitions and the peak maxima in the XAS data were used to determine the photon energies for the emission measurements. As observed in the RIXS data, Cr metal has an intense peak at ~3 eV below $E_F$ due to metal bonding with a bandwidth of 4 eV that is narrower than for $Cr_{0.51}N_{0.49}$. At 583.8 and non-resonant 625 eV excitation energy, Cr thin film shows $L_3$ and $L_2$ features due to spin-orbit split of Cr *3d* states contrary to $Cr_{0.51}N_{0.49}$. Due to hybridization of N *2p* states with Cr *3d* states, additional crystal field splitting $t_{2g}$-$e_g$ takes place around the $L_3$ and $L_2$ features in $Cr_{0.51}N_{0.49}$.